\documentclass[twoside,twocolumn,9pt]{article}
\usepackage{extsizes}
\usepackage[super,sort&compress,comma]{natbib} 
\usepackage[version=3]{mhchem}
\usepackage[left=1.5cm, right=1.5cm, top=1.785cm, bottom=2.0cm]{geometry}
\usepackage{balance}
\usepackage{times,mathptmx}
\usepackage{sectsty,amssymb,amsmath,wasysym}
\usepackage{graphicx} 
\usepackage{lastpage}
\usepackage[format=plain,justification=justified,singlelinecheck=false,font={stretch=1.125,small,sf},labelfont=bf,labelsep=space]{caption}
\usepackage{float}
\usepackage{fancyhdr}
\usepackage{fnpos}
\usepackage[english]{babel}
\addto{\captionsenglish}{%
	
}
\usepackage{array}
\usepackage{droidsans}
\usepackage{charter}
\usepackage[T1]{fontenc}
\usepackage[usenames,dvipsnames]{xcolor}
\usepackage{setspace}
\usepackage[compact]{titlesec}
\usepackage{tikz,pgfplots}
\usepackage{bm}
\usepackage[hidelinks=true]{hyperref} 
\hypersetup{
	colorlinks   = true, 
	urlcolor     = blue, 
	linkcolor    = blue, 
	citecolor   = red 
}
\usepackage{stackengine}
\usepackage{pgfplotstable}
\usetikzlibrary{arrows}
\usepgfplotslibrary{fillbetween}
\pgfplotsset{compat=1.10}

\usepackage{relsize}

\tikzset{
	font={\fontsize{8pt}{10}\selectfont}}

\definecolor{blues1}{RGB}{198, 219, 239}
\definecolor{blues2}{RGB}{158, 202, 225}
\definecolor{blues3}{RGB}{107, 174, 214}
\definecolor{green3}{RGB}{107, 174, 21}
\definecolor{blues4}{RGB}{49, 130, 189}
\definecolor{green4}{RGB}{49, 130, 18}
\definecolor{blues5}{RGB}{8, 81, 156}
\usepackage{tikz}
\usetikzlibrary{arrows}

\pgfplotscreateplotcyclelist{colorbrewer-blues}{
	{blues1},
	{blues2},
	{blues3},
	{blues4},
	{blues5},
}

\definecolor{cream}{RGB}{222,217,201}

\def\k{\kappa}

\def\f12{\frac{1}{2}}

\def\m{\rm{m}}

\def\m{m}

\def\ga{\gamma}

\def\k{\kappa}
\def\g{\gamma}

\def\lec{l_\text{ec}}

\def\ty{\vartheta_Y}
\def\O{\mathcal{O}}
\def\bb{\epsilon^{-1}}

\def\ndt{\tilde{T}}

\def\um{\mu \text{m}}
\def\mm{\text{mm}}
\def\nm{\text{nm}}
\def\mN{\text{mN}}
\def\uN{\mu N}
\def\m{\text{m}}
\def\mW{\text{mW}}
\def\m{\text{m}}
\def\cm{\text{cm}}
\def\kPa{\text{kPa}}
\def\MPa{\text{MPa}}

\def\eb{\text{eb}}

\begin{document}
	
	\pagestyle{fancy}
	\thispagestyle{plain}
	\fancypagestyle{plain}{
		
		\fancyhead[C]{\includegraphics[width=18.5cm]{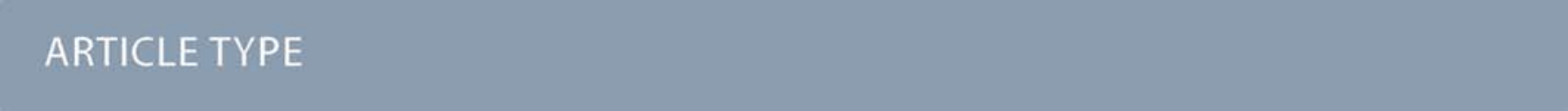}}
		\fancyhead[L]{\hspace{0cm}\vspace{1.5cm}\includegraphics[height=30pt]{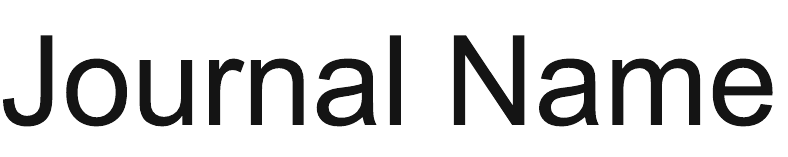}}
		\fancyhead[R]{\hspace{0cm}\vspace{1.7cm}\includegraphics[height=55pt]{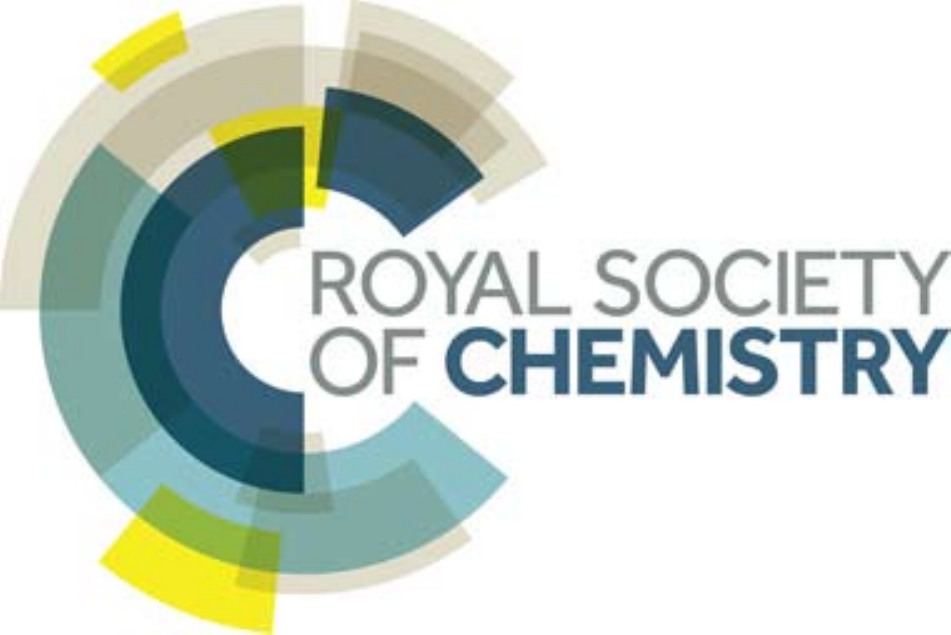}}
		\renewcommand{\headrulewidth}{0pt}
	}
	
	\makeFNbottom
	\makeatletter
	\renewcommand\LARGE{\@setfontsize\LARGE{15pt}{17}}
	\renewcommand\Large{\@setfontsize\Large{12pt}{14}}
	\renewcommand\large{\@setfontsize\large{10pt}{12}}
	\renewcommand\footnotesize{\@setfontsize\footnotesize{7pt}{10}}
	\makeatother
	
	\renewcommand{\thefootnote}{\fnsymbol{footnote}}
	\renewcommand\footnoterule{\vspace*{1pt}%
		\color{cream}\hrule width 3.5in height 0.4pt \color{black}\vspace*{5pt}} 
	\setcounter{secnumdepth}{5}
	
	\makeatletter 
	\renewcommand\@biblabel[1]{#1}            
	\renewcommand\@makefntext[1]%
	{\noindent\makebox[0pt][r]{\@thefnmark\,}#1}
	\makeatother 
	\renewcommand{\figurename}{\small{Fig.}~}
	\sectionfont{\sffamily\Large}
	\subsectionfont{\normalsize}
	\subsubsectionfont{\bf}
	\setstretch{1.125} 
	\setlength{\skip\footins}{0.8cm}
	\setlength{\footnotesep}{0.25cm}
	\setlength{\jot}{10pt}
	\titlespacing*{\section}{0pt}{4pt}{4pt}
	\titlespacing*{\subsection}{0pt}{15pt}{1pt}
	
	\fancyfoot{}
	\fancyfoot[LO,RE]{\vspace{-7.1pt}\includegraphics[height=9pt]{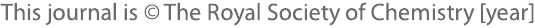}}
	\fancyfoot[CO]{\vspace{-7.1pt}\hspace{13.2cm}\includegraphics{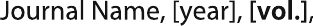}}
	\fancyfoot[CE]{\vspace{-7.2pt}\hspace{-14.2cm}\includegraphics{head_foot/RF}}
	\fancyfoot[RO]{\footnotesize{\sffamily{1--\pageref{LastPage} ~\textbar  \hspace{2pt}\thepage}}}
	\fancyfoot[LE]{\footnotesize{\sffamily{\thepage~\textbar\hspace{3.45cm} 1--\pageref{LastPage}}}}
	\fancyhead{}
	\renewcommand{\headrulewidth}{0pt} 
	\renewcommand{\footrulewidth}{0pt}
	\setlength{\arrayrulewidth}{1pt}
	\setlength{\columnsep}{6.5mm}
	\setlength\bibsep{1pt}
	
	\makeatletter 
	\newlength{\figrulesep} 
	\setlength{\figrulesep}{0.5\textfloatsep} 
	
	\newcommand{\topfigrule}{\vspace*{-1pt}%
		\noindent{\color{cream}\rule[-\figrulesep]{\columnwidth}{1.5pt}} }
	
	\newcommand{\botfigrule}{\vspace*{-2pt}%
		\noindent{\color{cream}\rule[\figrulesep]{\columnwidth}{1.5pt}} }
	
	\newcommand{\dblfigrule}{\vspace*{-1pt}%
		\noindent{\color{cream}\rule[-\figrulesep]{\textwidth}{1.5pt}} }
	
	\makeatother
	
	\twocolumn[
	\begin{@twocolumnfalse}
		\vspace{3cm}
		\sffamily
		\begin{tabular}{m{4.5cm} p{13.5cm} }
			
			\includegraphics{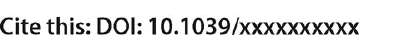} & \noindent\LARGE{\textbf{Wetting and wrapping of a floating droplet by a thin elastic filament}} \\
			\vspace{0.3cm} & \vspace{0.3cm} \\
			
			& \noindent\large{S Ganga Prasath,$^\dag$\textit{$^{a,b}$} Joel Marthelot,\textit{$^{c}$}, Narayanan Menon\textit{$^{d}$} and Rama Govindarajan,\textit{$^{a}$}} \\
			
			\includegraphics{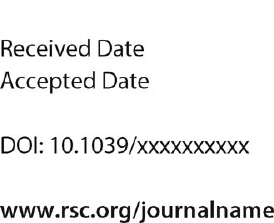} & \noindent\normalsize{We study the wetting of a thin elastic filament floating on a fluid surface by a droplet of another, immiscible fluid. This quasi-2D experimental system is the lower-dimensional counterpart of the wetting and wrapping of a droplet by an elastic sheet. The simplicity of this system allows us to study the phenomenology of partial wetting and wrapping of the droplet by measuring angles of contact as a function of the elasticity of the filament, the applied tension and the curvature of the droplet. We find that a purely geometric theory gives a good description of the mechanical equilibria in the system. The estimates of applied tension and tension in the filament obey an elastic version of the Young-Laplace-Dupr\'e relation. However, curvatures close to the contact line are not captured by the geometric theory, possibly because of 3D effects at the contact line. We also find that when a highly-bendable filament completely wraps the droplet, there is continuity of curvature at the droplet-filament interface, leading to  seamless wrapping as observed in a 3D droplet.  } \\			
		\end{tabular}
		
	\end{@twocolumnfalse} \vspace{0.6cm}
	
	]
	
	\renewcommand*\rmdefault{bch}\normalfont\upshape
	\rmfamily
	\section*{}
	\vspace{-1cm}

	
	\footnotetext{\textit{$^{a}$~International Centre for Theoretical Sciences (ICTS-TIFR) Shivakote, Hesaraghatta Hobli, Bengaluru 560089, India.}}
	\footnotetext{\textit{$^{b}$~School of Engineering and Applied Sciences, Harvard University, Cambridge, MA 02143, USA.}}
	\footnotetext{\textit{$^{c}$~Aix-Marseille University, CNRS, IUSTI (Institut Universitaire des Syst\'emes Thermiques Industriels), 13013 Marseille, France.}}
	\footnotetext{\textit{$^{d}$~Department of Physics, University of Massachusetts Amherst, Amherst, MA 01003, USA.}}
	
	\footnotetext{\dag~E-mail: gangaprasath@seas.harvard.edu}

	\section{Introduction}
	The wetting properties of a drop on a rigid substrate determines the angle of contact with the solid, known as the Young angle of contact. However if the substrate is a soft solid\cite{jerison2011,style2013,karpitschka2016,pandey2017} or an unstretchable but thin film\cite{huang2007,schroll2013,toga2013}, it can deform under the capillary action of the drop. In the former case, the liquid-vapour surface tension induces large localised stretching close to the contact line while in the latter, the film bends without stretching, resulting in large bending close to the contact-line. In both these scenarios the perceived contact angle at scales of the droplet-size deviates from the Young's contact angle, $\vartheta_Y$~\cite{andreotti2016}. The magnitude of this contact angle is obtained from a global energy minimisation rather than a  simple local force balance~\cite{schroll2013,brau2018}. This anomalous contact angle behaviour has several applications ranging from bio-locomotion to creating hydrophobic fabric~\cite{bico2004,chandra2009,duprat2012,bush2006}. 
	
	\begin{figure*}
		\centering
		\includegraphics[scale=0.8]{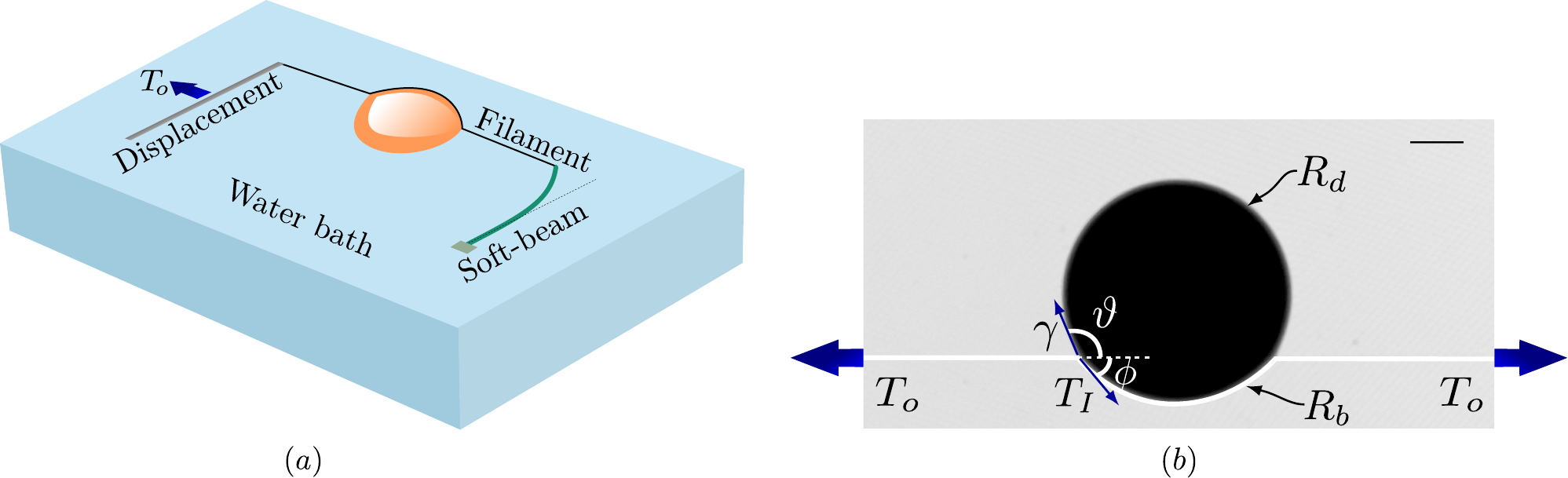}
		\caption{$(a)$ Schematic of the experimental setup where a droplet of Mineral oil (orange) is placed in the vicinity of a floating thin elastic filament at the air-water interface of a water bath. One end of the filament is connected to a soft beam (green) whose end displacement is used to measure the applied tension $T_o$. Tension in our experiments is controlled by inducing in-plane displacement of the free end of the filament. $(b)$ Variables of interest are shown on top of an image from experiments : $\vartheta, \phi$ - angle made by the droplet with the buckled filament; $T_o$ - applied boundary tension and $T_I$ - tension in the droplet-wet region, $\g$ - effective surface tension of the 2-D droplet along `contact-line' as detailed in the article; $R_d, R_b$ - radius of curvature of the free interface of droplet and the radius of curvature of the filament wet by droplet. The scale bar in the image is $5\mm$.}
		\label{fig:fig1}
	\end{figure*}
	We study a two-dimensional version of the wetting experiment (as shown in fig.~\ref{fig:fig1}$(a)$) at an air-water interface, where the thin sheet is replaced by a slender elastic filament and is wet by a nearly flat oil drop floating on the water interface.  The three-dimensional problem of a droplet of liquid on top of a floating thin sheet was initially studied because the capillary forces at the contact-line generate a radial wrinkling pattern~\cite{huang2007,schroll2013}. However, from the viewpoint of studying the contact angle, the wrinkles are a hindrance, as they impede the measurement of the contact angle and the deformation of the sheet close to the contact line where there is large localised bending. The 2D system we consider here does not have the complications of wrinkles. In the regime of wetting geometry where droplet size approaches the size of the sheet, if the sheet is highly bendable then it can wrap around the droplet and enclose it entirely~\cite{paulsen2015,deepak2018}. Wrapping in 2D can occur with smooth, isometric bending, unlike in 3D, thus the 2D filament-droplet system allows us to study phenomenon of wrapping of the droplet by the filament more easily than in the 3D system.
	
	We obtain the entire range of phenomena, from the Young contact geometry, to large deviations from apparent Young contact, to wrapping by two methods.  The first of these is to tune the competition between the liquid-vapour surface tension, and the bending rigidity of the sheet over the scale of the drop. This competition is captured in a dimensionless parameter called the bendability, which is the ratio of the droplet-size, $w$ to the capillary-bending length, $\lec$ which is the length scale at which bending and capillary forces are similar in magnitude. Most studies~\cite{py2007,antkowiak2011,rivetti2012,pineirua2010,elettro2016,neukirch2013} are in the regime where the bendability is $\O(1)$ while our experiments are in the high-bendability limit. We achieve this limit by using thin filaments and large droplet sizes. This guarantees a separation in scales between droplet size and capillary-bending length, and the behaviour in this limit is dramatically different from that in the low bendability limit~\cite{schroll2013,paulsen2015,deepak2018}. The second method we use to traverse these phases is to smoothly modify the rigidity of the filament by varying the tension imposed on the filament. As we will show in this article, when the filament is under large tension, it does not deform under the capillary force of the droplet, and Young scenario is recovered; when it is slack, then large deformation and wrapping can occur.
	
	In the absence of tension, the sheet can undergo a budding transition explored in the context of lipid membranes~\cite{kusumaatmaja2011droplet,dimova2012lipid}. When a large droplet of liquid is in contact with a lipid membrane, the bending energy of the membrane can be neglected and in this limit the membrane can change its shape to wrap the droplet completely. This transition is driven by competition between membrane tension and the wetting property of the droplet, analogous to the wrapping transition of an elastic sheet~\cite{brau2018}. On the other hand when an elastic nanoparticle or a vesicle~\cite{yi2011cellular,yi2014phase} is covered by a lipid membrane, deformations of the vesicle can now cost energy. When the adhesion energy between the membrane and vesicle as well as the surface tension of the exposed vesicle are varied, they can exhibit three phases: no wrapping, partial wrapping, full-wrapping. Interestingly, in the limit of vanishing bending stiffness of the vesicle, the theory of Yu et al.~\cite{yi2011cellular,yi2014phase} coincides with the calculation of Brau et al.~\cite{brau2018} for a 2-D elastic filament wrapping a macroscopic droplet. In addition, Brau et al.~\cite{brau2018} provide predictions for the effects of finite tension, which we explore in this article.
	
	As shown in fig.~\ref{fig:fig1}$(b)$, we measure the contact angles $\vartheta, \phi$ of a partially wetted filament as a function of applied tension.  We show in this paper that the contact angles exhibit a universal behaviour in the thin-filament limit which were not measured in earlier work on the equivalent 3D systems\cite{toga2013,huang2007,schroll2013}. In the limit of infinite bendability, our results match well with a recent theory~\cite{brau2018}, with the effective surface tension of the 2-D droplet being the only fitting parameter. Under large magnitudes of applied tension, the theory predicts that though $\phi$ and $(\ty-\vartheta) \rightarrow 0$, the ratio $(\ty - \vartheta)/\phi$ asymptotes to 1/2. We measure this ratio in our experiments and observe the trend predicted by the theory to hold true. We also find that the applied tension and the tension in the buckled zone obey a force balance relation remarkably similar to the Young-Laplace relation. This relation is shown by Brau et al.~\cite{brau2018} to arise out of global energy minimisation. In the theory of vanishing bending stiffness, the region close to the contact-line is of infinite curvature. However, in the experiments the filament has a finite magnitude of curvature denoted by $1/R_{\eb}$, an elasto-bending scale; we study the variation of $R_{\eb}$ with $T_o$, the applied boundary tension. The effects of finite bendability here produce deviations from the high bendability theory. In the wrapped regime when the filament length is less than the perimeter of the droplet, we find that the filament forms part of a circle around the drop, unlike a 3D wrapping experiment\cite{paulsen2015} where the optimal wrapping assumes the shape of a parachute or a mylar balloon resembles a parachute. The circular shape ensures that there is no jump in curvature at the interface between filament end and droplet interface; this leads to seamless wrapping as observed in the case of ultra-thin sheets~\cite{deepak2018}. On the other hand, for low-bendability filaments, the radius of curvature of the droplet-interface diverges as the ends of the filament approach each other.
	\section{Relevant variables}\label{sec:relvar}
	Four length scales govern the overall mechanics of the filament-drop system. These are the diameter $d$ of the filament, the droplet radius $w$, the length of the filament $L$, the capillary-bending length scale $\lec = (B/\ga)^{1/3}$ with $\g$ being the droplet-vapour surface tension, $B$ the bending stiffness,  given for a filament of circular cross-section as: $B=E\pi d^4/64$, ($E$ - Youngs' modulus of filament material). A fifth length scale arising out of stretching, $l_m = \g/E$  is irrelevant in our experiments because $l_m = 0.03\um \ll d \approx \O(10 \um)$.  Moreover we are interested in a high-bendability regime where the size of droplet is much larger than the capillary-bending length scale, $w \gg \lec$. The non-dimensional quantity that indicates this scale separation is the bendability: $\bb = (w/\lec)^2$.
		\begin{figure*}
		\centering
		\includegraphics[scale=1.25]{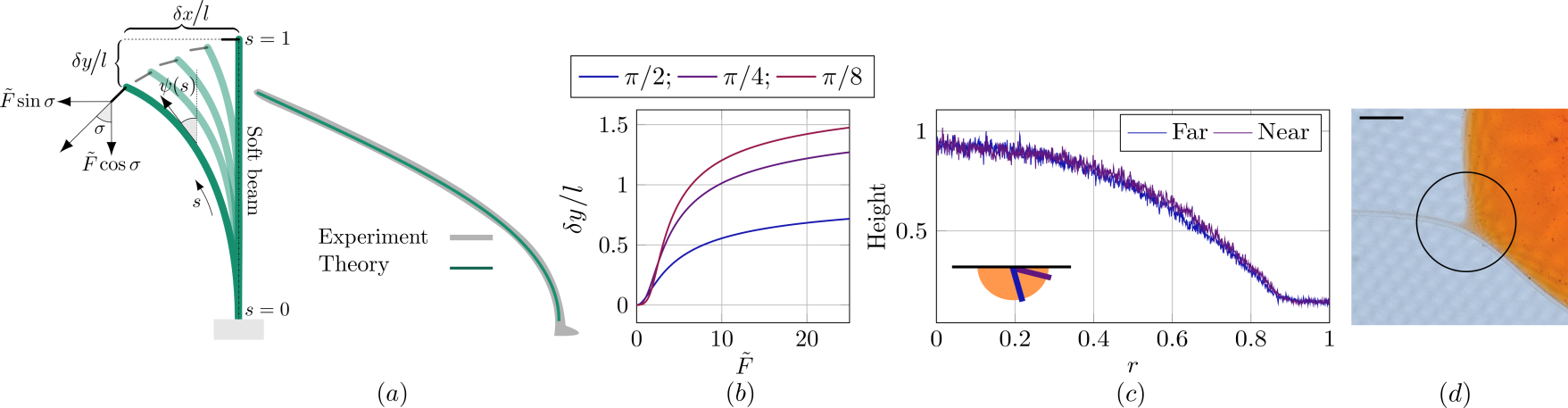}
		\caption{$(a)$ Schematic of the soft beam attached to one end of the filament to measure tension in the filament with the relevant variables: $\psi(s)$ - angle between tangent and vertical; $\tilde{F}$ - non-dimensional  force applied at beam's end; $\sigma$ - angle at which the force is applied; $s$ - non-dimensional arc-length; $\delta x/l, \delta y/l$ - non-dimensional displacement along $x, y$-direction. Alongside is the  image of the beam from the experiment superimposed on the numerically-solved shape from eqn.~\ref{eq:beam}. $(b)$ Numerically computed displacement $\delta y/l$ vs non-dimensional force, $\tilde{F}$ for three different $\sigma$ values: $\pi/2, \pi/4, \pi/8$. $(c)$ Fluorescence intensity along two radial lines, one close to the contact-line of PDMS filament and one far from it (shown schematically in the inset).  This indicates that the presence of the filament does not distort the 3D height profile of the drop, except perhaps very close to the 4-phase contact line as shown in $(d)$ where we see 3-D effects appear as the droplet becomes thin near the contact line, seen as a region of wet zone with large changes in interface curvature. The scale bar in the image is $2.5\mm$.}
		\label{fig:fig2}
	\end{figure*}

	By varying these length-scales and the applied tension we can explore the phases of the filament-drop phase diagram. The axes of the diagram are three non-dimensional quantities constructed from these variables:
	\begin{equation}
	\ndt = {T_o/(\g d)}, \quad
	\bb = (w/\lec)^2, \quad
	\Phi = (L/w).
	\end{equation}
	Here $T_o$ is the applied tension at the boundary, as seen in fig.~\ref{fig:fig1}$(b)$; $\ndt$ is the non-dimensional applied tension; $\Phi$ the ratio of filament length $L$ to droplet radius $w$. In the partially wet regime we maintain the limit: $\bb \gg 1, \Phi \gg 2\pi$. In the second part of the article we explore the other regime of $\bb \gg 1, \Phi \leq 2\pi$ to understand the wrapping mechanism.
	
	\section{Experimental set-up}
	
	Our experiment consists of a thin elastic filament floating at air-water interface, placed in contact with a floating oil droplet as shown in fig.~\ref{fig:fig1}$(a)$. One end of the filament is attached to a translation stage and the other end to a beam made out of a soft elastic material (Vinyl polysiloxane). We control the tension in the filament by moving the beam. As described in detail later, the floating droplet is flattened by gravity and so behaves approximately like a two-dimensional object. We define the contact angles $\vartheta$ and $\phi$ in the plane of the air-water interface (see fig.~\ref{fig:fig1}$(b)$). The contact angle of the droplet is a scale dependent quantity and here we measure them at the scale of the size of the droplet. The effective surface tension of this 2-D droplet is the line tension of the droplet at the air-water interface. In the partially-wrapped state we vary the size of the droplet, diameter of the filament and the applied boundary tension, while in the wrapping experiments, we vary the droplet size for various filament diameters. These parameters help us explore different high-bendability morphologies of the filament. We image the filament shape and measure the curvature of the filament close to contact-line of the droplet.  Near full wrapping, we use fluorescence imaging to accurately capture the shape of the filament. This allows us to calculate the radius of curvature of the filament and that of the droplet as a function of its bendability.
	
	\subsubsection*{Making filaments}	
	The thin filaments used in our experiments are made out of Polydimethylsiloxane (PDMS) whose Young's modulus is $E=1\MPa$. We use a mixture of PDMS base (Sylgard 184, Dow Corning), accelerator and cross-linker in the ratio (10:2:1) at room temperature, and as the mixture begins to set, we take a droplet of this mixture and pull it using tweezers to create thin long threads which then set.  We make filaments whose diameter $d$ varies between $80\um - 200\um$ and with length, $L$ = $1000d$.
	
	\subsubsection*{Measuring tension}
	The capillary forces originate from surface tensions that are~$\sim~10 \mN/\m$, and they act  on filaments $\sim 100 \um$ in diameter, resulting in forces of $\O(\uN)$ in magnitude; this demands a sensitive force sensor. We clamp a long, soft beam at one end to a translation stage and attach the filament at its other end.  This is shown schematically in fig.~\ref{fig:fig1}$(a)$. The soft beam is made of Vinyl polysiloxane (VPS) and has a diameter of $0.5 \mm$, length of $l=6 \cm$ and Young's modulus of $E_s=200\kPa$. When the translational stage is displaced the beam bends and we track the tip of the soft beam at the end attached to the filament. The standard small deflection approximation gives the dependence of force on displacement as: $F \approx 3 B_s \delta x / l^3$ where $B_s$ is the bending stiffness of soft beam, given by $B_s=E_s \pi t^4/4$, $t$~-~beam diameter, $\delta x$~-~horizontal displacement, $l$~-~length of soft-beam as shown in fig.~\ref{fig:fig2}$(a)$.

	\begin{figure*}
	\centering
	\includegraphics[scale=1.1]{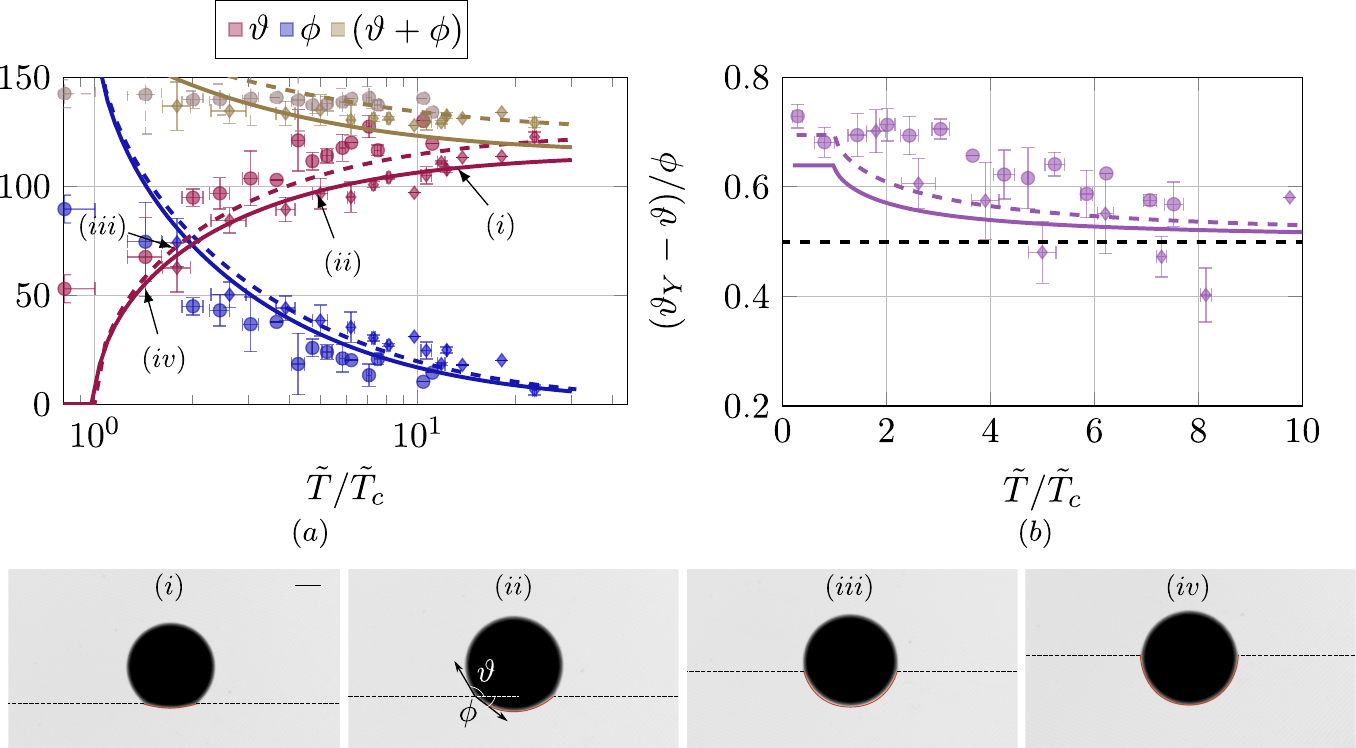}
	\caption{$(a)$ Measured values of $\vartheta$ and $\phi$ from experiments for different bendability values of the drop-filament system. For the filament diameter $205 \mu \m$, we choose four different droplet sizes: $12.7 \mm, 13.8 \mm, 18.0 \mm, 20.2 \mm$ shown using $\Circle$. Similar angles $\vartheta$ and $\phi$ measured for a filament of diameter $77 \um$ and three different droplet sizes : $10.6 \mm, 14.2 \mm, 16.6 \mm$ plotted as $\Diamond$. We see a clear collapse of all the data indicating a universal behaviour of perceived contact angle in the high-bendability limit, with solid line being eqns.~\ref{eq:theta},~\ref{eq:phi} for $\ty = 115^\circ$ (corresponding to the thick filament) and dashed line for $\ty = 125^\circ$. The point where $\vartheta, \phi$ intersect is where $\ndt=2\ndt_c$, which in our experiments give $T_c/d = 2.2 \mN/\m$, from $\gamma = 7.6 \mN/\m, \ty = 115^\circ$. $(b)$ Computed values of $(\ty-\vartheta)/\phi$ in experiments (symbols) compared to the theoretical predictions from eqns.~\ref{eq:theta},~\ref{eq:phi} which approach a value of $1/2$ at large tensions. $\Circle$ corresponds to the thicker and $\Diamond$ to the thinner of the filaments. $(i)-(iv)$ show the shape of filament-drop system as the tension in the filament is decreased. Scale bar in $(i)$ is $5 \mm$.}
	\label{fig:fig3}
	\end{figure*}	
	
	However, for the range of tension we are interested in, we require the solution to the full non-linear beam equation. Here, unlike in the small displacement limit, the displacement in the vertical direction of the beam, $\delta y$ cannot be neglected and for large tensions the displacement in this direction is sensitive to applied tension as shown in fig.~\ref{fig:fig2}$(b)$. Furthermore, the angle between the end of the filament and the initial configuration changes as a function of tension creating an angle $\sigma$. We measure $\sigma$, $\delta x$, $\delta y$ for each tension value in the experiments. We then find solutions to the non-linear problem given by:
	\begin{align}
	\psi''(s) + \tilde{F} \cos \sigma \sin \psi(s) + \tilde{F} \sin \sigma \cos \psi(s) =& 0, \label{eq:beam}
	\end{align}
	where $\psi$ is the angle between the tangent to the beam and vertical (see fig.\ref{fig:fig2}$(a)$), $s$ the non-dimensional arc-length along beam (non-dimensionalised using $l$) with $s=0$ indicating the fixed end and $s=1$ is the end connected to the elastic filament. $\tilde{F}=Fl^2/B_s$ is the non-dimensional applied force. We solve this system numerically using a shooting method~\cite{basilelec} under the boundary conditions: $\psi (0) = 0$, and $\psi'(1) = 0$. From the solution for different values of $\tilde{F}$ for a given $\sigma$ we look for the $\tilde{F}$ that corresponds to the measured $\delta x/l$ and $\delta y/l$, which is the required quantity to compute $T_o$. This procedure is executed for all displacements and we show in fig.~\ref{fig:fig2}$(a)$ that the computed shape (green line) matches well with the experimentally observed shape (gray line).

	\subsubsection*{Droplet and scale separation}
	We use mineral oil of density 0.86 g/mL for the droplets. The oil is dyed with Sudan red G, a hydrophobic, water-insoluble dye. This captures the shape of the droplet precisely as we image it under uniform light. The oil droplet is a three-dimensional object with three length-scales relevant at different regions of the droplet. These are $w$~-~radius of droplet, $d$~-~diameter of filament, $l_c$~-~capillary length. The droplet size $w$ is the largest of these length-scales, $d$ is relevant close to the region where the droplet is in contact with filament and $l_c$ in the curved region approaching air-water interface. For large droplets, gravity ensures a uniform thickness~$\sim l_c$ and if we work with small-$d$ filaments, then we are in a scale separated regime given by: $d \ll l_c \ll w$. Furthermore the quantities of interest in our experiments $\vartheta$ and $\phi$, are the macroscopic angles and not the microscopic wetting angle.  We ensure this separation by choosing $w$ in the range $0.5 \cm - 2 \cm$ with the capillary length of Mineral oil $l_c = 1.8 \mm$ and $d=80 \um-200\um$. In fig.~\ref{fig:fig2}$(c)$ we plot the droplet surface profile close to filament contact line, and far from it, to show that the shape of the droplet is not strongly perturbed in the thin direction by the presence of the filament and thus the 2-D approximation is valid.\footnote{We do not explore here in detail the question of why the filament chooses to sit at the three phase contact line instead of passing undeviated over or under the drop. Possibly the bending energy lost in curving around the drop is more than compensated by the oil-air and water-air interface protected by the filament.}
	
	\subsubsection*{Visualising filament}
	In order to visualise the filaments in our wrapping experiments we wet the filament in a solution of Nile Red in ethanol and allow the ethanol to evaporate. We leave the droplet uncoloured and shine a laser beam (\textsc{Wicked Laser}, $500\mW$ at $500\nm$) on the filament. The filament fluoresces in the red, and a filter is used to eliminate the green illumination line. Though Nile Red is a hydrophobic dye and diffuses into mineral oil at long timescale, our experiments were performed before the dye diffuses into the droplet.
	
	\subsection*{Highlights of theory}
	We make comparisons to a theory of a 2-D drop-on-sheet problem  $i.e.,$ a drop modelled as a cylinder sitting on top of an inextensible rectangular sheet, in the infinite-bendability limit~\cite{brau2018}. This involves minimising the total surface energy of the system, with contributions from liquid-vapour, liquid-solid, and solid-vapour interfaces. The analysis reduces to a purely geometric question with contact angle, $\ty$ and applied tension, $\ndt$ being the relevant parameters. Given these parameters, the complete shape of the droplet and sheet is predicted, for a fixed area of droplet. The expressions for the angles $\vartheta, \phi$ are given by~\cite{brau2018}:
	\begin{align}
	\cos \vartheta =& \ \frac{(1 + 2\ndt \cos \ty - \cos^2 \ty)}{2 \ndt}, \label{eq:theta} \\
	\cos \phi =& \ \frac{(1 - \cos \ty /\ndt + (\cos^2 \ty -1)/ 2\ndt^2)}{1 - \cos \ty / \ndt}. \label{eq:phi}
	\end{align}
	The critical tension is determined from the criterion for the validity of the above expressions i.e. $0 \leq \vartheta,\phi \leq \pi$ from which we get $\ndt_c = \cos^2(\ty/2)$. This is precisely the transition from a partially wet state where only a part of the drop boundary is covered by the sheet, to a wrapped state where the droplet boundary is completely encapsulated. Interestingly the tension at which $\vartheta=\phi$ occurs at $\ndt=2\ndt_c$.  We will use the critical tension, $\ndt_c$ to measure the unknown line tension $\gamma$ in our experiments.
	
	\section{Results}
	\begin{figure}
	\centering
	\includegraphics[scale=0.8,trim={60 0 60 0}]{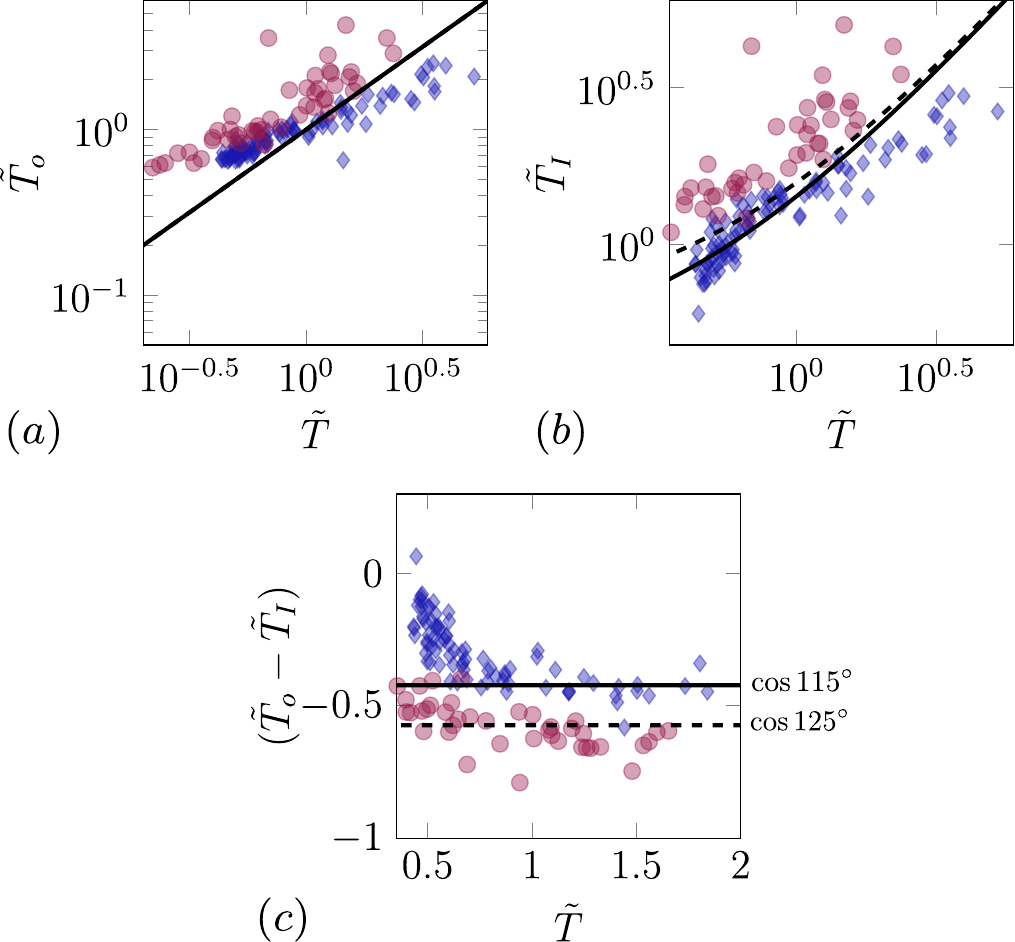}
	\caption{$(a)$ Non-dimensional applied tension $\ndt_o$, $(b)$ tension in the portion of the filament wet by the droplet $\ndt_I$, and $(c)$ the difference between these tension values, $(\ndt_o-\ndt_I)$ evaluated by using angles $\vartheta, \phi$ measured from experiments (using eqn.~\ref{eq:thetaphi}) as functions of the applied tension, $\ndt$ measured using soft-beam displacement. $\Circle$ correspond to filament diameter $205 \um$ and $\Diamond$ to diameter $77 \um$ for different droplet sizes. The solid and dashed lines respectively indicate theoretical predictions from eqn.~\ref{eq:tension}.}
	\label{fig:fig4}
	\end{figure}	
	\subsection*{Contact angle and critical wrapping tension}
	In fig.~\ref{fig:fig3}$(a)$ we plot  $\vartheta, \phi$ as functions of $\ndt$, scaled by the critical tension for wrapping, $\ndt_c$. We show data for filaments of two different diameters, $77 \um$ and $205 \um$. The data were taken by varying the tension while holding the drop size fixed. The drop size was then varied in the range $10.6 \mm - 20.2 \mm$. These two filament diameters and the different droplet sizes  helped us span bendability values between $\bb = 80-1920$.  In order to compute the bendability $\bb = (w/\lec)^2$ of the droplet-filament system, we need the line-tension of the oil droplet.  	
	To obtain $\gamma$, we note in fig.~\ref{fig:fig3}$(a)$ that as the magnitude of tension increases there is a crossover in magnitude between $\phi$ and $\vartheta$. From the infinite bendability theory we expect this cross-over to happen at $\ndt/\ndt_c=2$ (see eqns.~\ref{eq:theta} \& \ref{eq:phi}).  In order to match this cross-over point in experiments, $\gamma$ is the only fitting parameter. We find that  all the data collapse with the analytical expression in fig.~\ref{fig:fig3}$(a)$ for $\ga=7.6\mN /\m$. To extract $\ty$ we measure the contact angle for $\ndt \gg \ndt_c$. For this surface tension $\ga$, we find $T_c/d = 2.2 \mN /\m$ when $\ty=115^\circ$ for the thicker filament and $T_c/d = 1.6 \mN /\m$ when $\ty=125^\circ$ for the thinner filament.
	Now for $\ndt \gg \cos^2(\ty/2)$ we can expand the eqs.~\ref{eq:theta}, \ref{eq:phi} to get
	\begin{equation}
 	\phi \approx \frac{2 \sin \ty}{\ndt}, \ \ty - \vartheta \approx \frac{\sin \ty}{\ndt}.
	\label{eq:thetaphi}
	\end{equation}
	From this it is easy to see that for applied tension, $\ndt \gg 1$, the asymptotic value of the function $(\ty - \vartheta) / \phi \rightarrow 1/2$. Using the measured values of $\vartheta, \phi$ and $\ty$ in our experiments, we compute $(\ty - \vartheta)/\phi$ as shown in fig.~\ref{fig:fig3}$(b)$. The trend predicted by the analytical expression is captured, though the data are noisier at large tensions due to finite precision in the measurement of $\phi$ when its magnitude approaches zero.
	In the model of Brau et. al.~\cite{brau2018} the normalized internal stresses are related to geometric variables $\vartheta$ and $\phi$ alone through the expressions
	\begin{equation}
	\ndt_I = \frac{\sin \vartheta}{\sin \phi}, \ \ndt_o = \frac{\sin(\vartheta + \phi)}{\sin \phi}. \label{eq:tension}
	\end{equation}
	
	These expressions are derived by using the relation that the tension in the buckled zone is given by $T_I/(d R_b)=\gamma/R_d$ where $R_b, R_d$ are the radius of buckled zone of sheet and the radius of the droplet interface. Now the ratio $\ndt_I =T_I/(d \gamma) = R_b/R_d$ can be written in terms of $\vartheta, \phi$. The expression for $\ndt_o$ on the other hand is derived by minimising the total energy of the system as detailed in the previous section. In fig.~\ref{fig:fig4}$(a)$ we compare the boundary tension $\ndt_o$ computed from $\vartheta, \phi$ and eqn.~\ref{eq:tension} and the direct measurement of tension using the deflection of the soft beam-$\ndt$.  We next compare in fig.~\ref{fig:fig4}$(b)$ experimental data for $\ndt_I$,  the stress in the buckled zone obtained from the measured contact angles, to the analytical expression for $\ndt_I$ for two different $\ty=115^\circ, 125^\circ$ corresponding to the thick and thin filament. The jump in the magnitude of tension at the contact line is proportional to the equilibrium contact-angle $\ty$ and is given by:
	\begin{equation}
	(\ndt_o - \ndt_I) = \cos \ty, \label{eq:elYLD}
	\end{equation}
	which is the equivalent of the Young-Laplace-Dupr\'e relation for elastic filaments or sheets. In fig.~\ref{fig:fig4}$(c)$ we compare the difference $(\ndt_o-\ndt_I)$ as a function of applied tension $\ndt$ and see that it remains constant, close to the value $\cos \ty$, predicted by theory. There is a reasonable match between experiments and the above expression \ref{eq:tension} with the biggest deviations at small tensions. 

	\subsection*{Close to the contact-line}
	The angles and tensions measured in the previous section were compared with an infinite-bendability theoretical model. In this section, we explore quantities that reflect more obviously the finite bendability of the filaments. In the infinite-bendability prediction, the filament is buckled into a circle of radius $R_b$ where it is in contact with the droplet, and straight elsewhere with a sharp cusp connecting the straight region and the wetted region. However, finite bendability introduces finite curvature, $1/R_{\eb}$, in this transition zone close to the contact line.
	
	\begin{figure}[ht!]
	\centering
	\includegraphics[scale=0.85]{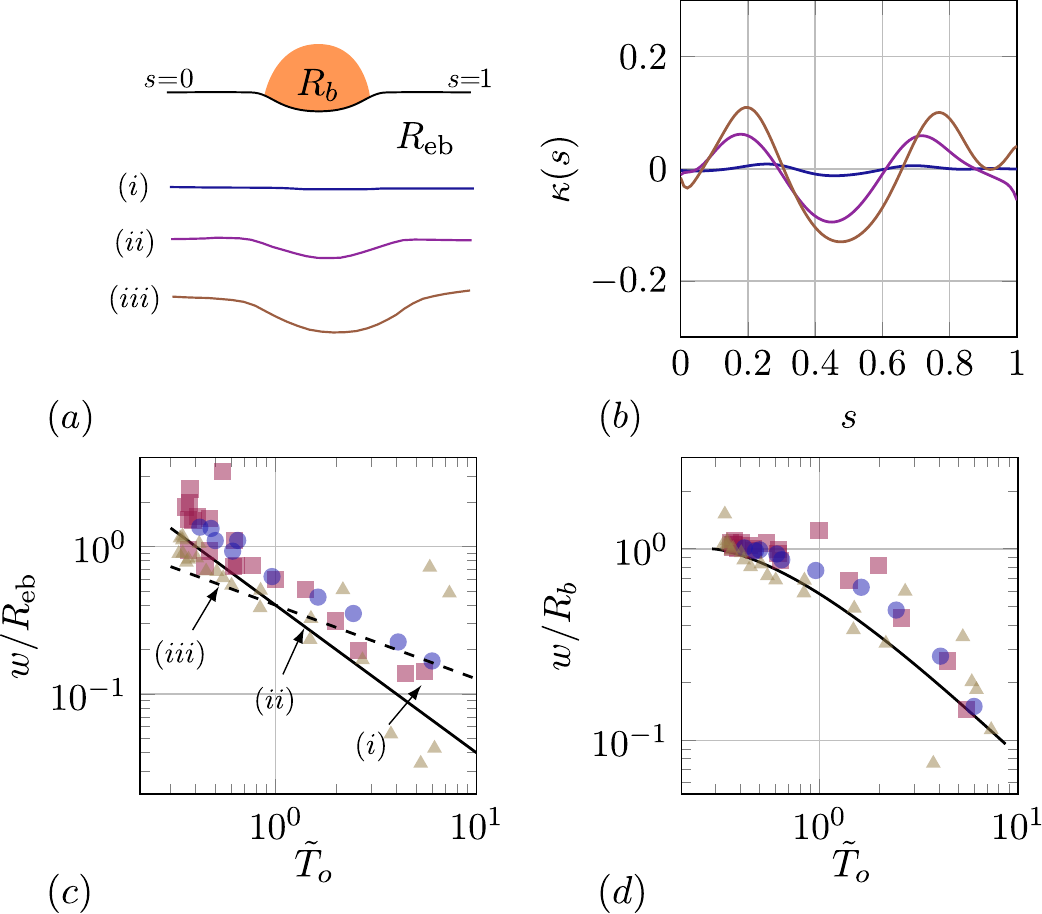}
	\caption{$(a)$ Three different filament shapes extracted from experiments as the applied tension $T_o$ is decreased for a fixed droplet size of $w=6.2 \mm$ and filament diameter $d=120 \um$. $(b)$ Corresponding signed curvature, $\k(s)$ of shapes in $(a)$ computed as a function of non-dimensional arc length $s$ (scaled using filament length) after fitting Bezier spline to the extracted shapes. $1/R_{\text{eb}}, 1/R_b$ correspond to maximum and minimum of $\k({s})$. $(c)$ The non-dimensional curvature of transition-zone close to the contact-line between filament and droplet, $w/R_{\text{eb}}$ and $(d)$ the non-dimensional curvature of the droplet wet part of filament, $w/R_b$ as a function of non-dimensional boundary tension $\ndt_o$ for a filament of diameter $d=120 \um$ for three different droplet sizes $w=6.2 \mm (\triangle), 7.6 \mm (\Circle), 9.4 \mm (\square)$ and several values of $T_o$, non-dimensionalised using an estimate of $\gamma = 5 \mN/\m$. The corresponding bendability values are $\bb = 49, 72, 125$. The solid line in $(c)$ corresponds to $\ndt_o^{-1}$ and the dashed line is $\ndt_o^{-1/2}$ while the solid line in $(d)$ corresponds to the complete expression for $\ndt_o$ using eq.~\ref{eq:area} for $\epsilon = 0$.}
	\label{fig:fig5}
	\end{figure}
	
	We measure $1/R_{\eb}$ as a function of applied tension, $\ndt$ for three different droplet sizes ($w = 6.2 \mm, 7.6 \mm, 9.4 \mm$) as shown in fig.~\ref{fig:fig5}$(c)$. To image the filament shape near the droplet, we dye the filament (and not the droplet as in the previous section) with Sudan Red G and illuminate with a uniform white light source.  After extracting the filament shape, we do a B-Spline curve fit to the filament shape and calculate the signed curvature as a function of filament arc length as shown in fig.~\ref{fig:fig5}$(b)$. The three shapes in fig.~\ref{fig:fig5}$(a)$ correspond to three different tensions in decreasing magnitude for a fixed droplet size of $w=6.2 \mm$. We identify the maximum value of curvature max($\k(s)$) with $1/R_\eb$ and the minimum value, min($\k(s)$) with $1/R_b$. We find that both curvatures decrease with increasing outer tension $T_o$ as shown in fig.~\ref{fig:fig5}$(c,d)$. 
	
	The curvature $1/R_b$ of the buckled part of the filament may be estimated in terms of the 2D Laplace pressure of the free surface of the droplet. We can relate $R_b$ as a function of $\ndt_o$ using the expression for area as
	\begin{equation}
	\pi w^2/R_b^2 = \frac{\sin^2 \phi ( \vartheta - \sin (2\theta) /2 ) + \sin^2 \vartheta ( \phi - \sin (2\phi)/2 ) }{\sin^2 \vartheta}, \label{eq:area}
	\end{equation}
	where $\vartheta, \phi$ are related to tension under eq.~\ref{eq:theta},~\ref{eq:phi}. The length-scale $R_\eb$ close to the contact line can be estimated from balancing bending forces and tension in the filament. In an idealized 2D situation, the tension jumps across the contact line resulting in a change in curvature from ${\epsilon^{-1/2} \ndt_I^{-1/2}}$ in the region wet by the droplet to ${\epsilon^{-1/2} \ndt_o^{-1/2}}$ outside. The experimental data for  $w/R_\eb$ in fig.~\ref{fig:fig5}$(c)$ do not clearly show the $\ndt_I^{-1/2}$ trend (dashed line), and are perhaps closer to $\ndt_o^{-1}$ (solid line). There are three major differences between the experiment and the model: one is that the filament bendability ranges from 74-190, whereas the model assumes large bendability; this leads to poor scale-separation between $R_\eb$ and $R_b$. A second respect in which the experiment is non-ideal is that the meniscus near the contact line is fully 3-dimensional, as shown in fig.~\ref{fig:fig2}$(d)$. Lastly the theoretical estimates also assume fixed area of the droplet while in the experiments a fixed volume of droplet is maintained as the tension is varied. In this process the projected area does vary, however there is only $5\%$ change in area over the range of tensions in the experiment. 
	
%
	\subsection*{Wrapping process}
\begin{figure}
		\centering
		\includegraphics[scale=0.8]{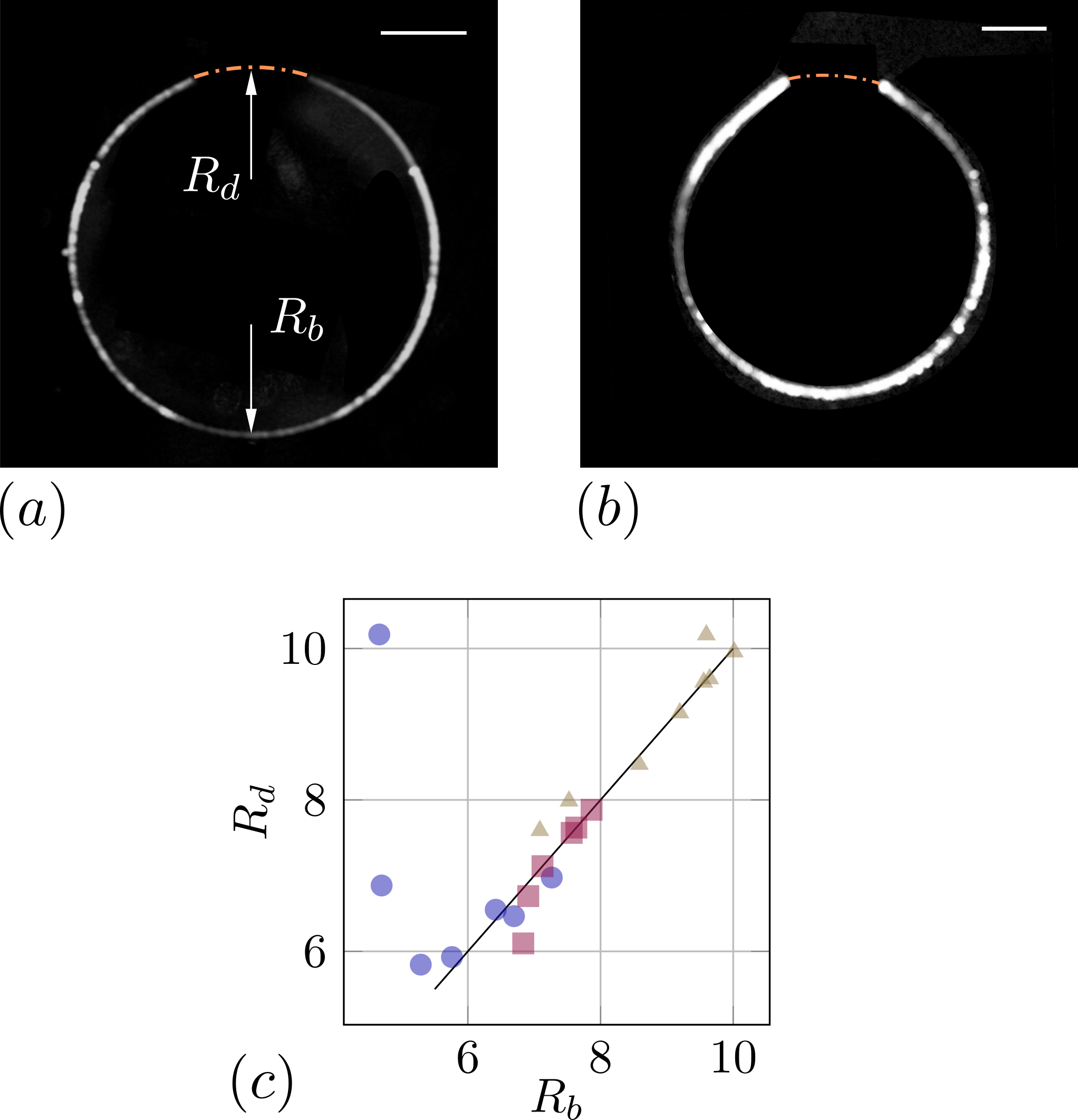}	
		\caption{$(a, b)$ Fluorescence images showing shape of filament encapsulating the droplet for a filament of diameter $d=90 \um, 170 \um$ and a droplet of size $w= 7.5 \mm, 2.3 \mm$ with droplet-interface shown as dashed-line. The scale bar is $3.7 \mm$. $(c)$ Radius of curvature of the buckled zone in the filament, $R_b$ vs the radius of curvature of free interface of the droplet, $R_d$ in $\mm$. Solid line indicates $R_b=R_d$. We consider three different filament diameters $d = 80 \um (\bigtriangleup), 90 \um (\Box), 170 \um (\Circle)$.}
		\label{fig:fig6}
	\end{figure}

	In order to emulate the process of complete wrapping of the droplet, we adopt the following procedure. A freely floating filament is brought into contact with a floating droplet such that $\Phi<2\pi$ and the droplet size is reduced until the ends of the filament come close to touching. We extract the shape of the filament from fluorescence images such as in fig.~\ref{fig:fig6}$(a)$. This procedure is followed for three different filament diameters, $d=80 \um, 90 \um, 170 \um$. The parameters $R_d$ and $R_b$, the radius of curvature of the droplet-water interface and radius of curvature of the buckled zone describe the geometry of the droplet-filament system. In the high-bendability limit, as we shall see, these are enough to describe the system's entire shape.
	
	First, we observe that the thinnest filament has a constant curvature along the length as shown in fig.~\ref{fig:fig6}$(a)$. This radius of curvature matches that of the droplet interface radius of curvature, $R_d$ plotted in fig.~\ref{fig:fig6}$(c)$ corresponding to open triangles ($\bigtriangleup$), where the solid line indicates $R_d=R_b$. As the droplet size decreases the buckling radius decreases as does the droplet interface radius. However the thickest filament corresponding to open circles ($\Circle$) shows a deviation from the straight line hinting that the interface becomes flatter before the ends come in contact. This divergence is a low-bendability effect which does not exist for the thinnest filament as seen in fig.~\ref{fig:fig6}$(b)$.
	
	Second, in the high bendability limit the shape of the filament is part of a circle which is in contrast with the behaviour of an axisymmetric sheet seen in Paulsen et al.~\cite{paulsen2015} where the solution is not part of a sphere but resembles that of parachute. This difference comes about from geometric constraints of area being preserved in inextensible sheets where it is the length that is preserved in filaments. From the force-balance equation we have:
	\begin{align}
	B \bigg[ \ddot{\k}(s) + \frac{\k^3(s)}{2}  \bigg] - T_o \k(s) &= - \frac{\ga}{w}
	\end{align}
	where $\k(s)$ is the curvature along the filament. Now in the high-bendability limit the dominant balance comes from tension in filament and droplet surface tension, $\ga$. However since $T_o \approx \gamma$ this implies $\k \approx w^{-1}$. The partially wrapped circular droplet now obeys the Young-Laplace relation over free interface of the droplet and the elastic Young-Laplace relation in eq.~\ref{eq:elYLD} over the filament. We see that $w$ determines the entire shape of the filament-droplet system, independent of any physical parameters in the system, just as in the unwrapping scenario. The seamless wrappings of drops within sheets seen in Kumar et al.~\cite{deepak2018} must be a consequence of the continuity of curvature at the sheet-droplet boundary, as we observe in the case of filaments. This is in stark contrast with the capillary origami~\cite{py2007}, where the wrapped state of the origami presents openings with pointed ends because $\bb \sim \O(1)$.

	\section{Conclusion}
	The two dimensional experimental system developed here to study capillary bending and wrapping illuminates some features of its three dimensional counterpart. We have shown that contact angles and critical wrapping tension of the filament can be described by an infinite bendability theoretical model~\cite{brau2018} and have also characterized the finite bendability effect close to the contact line. The tension inside the wetted region and the applied tension obey the predicted elastic version of the Young-Laplace-Dupr\'e relation. However, some features of the experimental system, such as the 3D geometry of the drop near the contact line, affect the comparison with the purely 2D model. In our experiments, the line tension of the droplet at the air water interface was a fitting parameter and the physical mechanism behind this effective surface tension need further probing. The 2-D system also provides a venue to further explore different phases in the phase-diagram of Brau et al.~\cite{brau2018} such as the partially wrapped and the completely wrapped phase. 
	
	\section{Acknowledgement}
	The authors wish to thank Benny Davidovitch and Fabian Brau for critical reading of the manuscript (and to Fabian in particular for pointing out issues with earlier version of fig. 5), Lee Walsh and Deepak Kumar for several insightful discussions. SGP would like to thank TCIS-TIFR where the experiments were initiated and the hospitality of people at UMass Amherst where part of the work was done. NM acknowledges support through NSF-DMR 1905698 and RG acknowledges support of the Department of Atomic Energy, Government of India, under project no. 12-R\&D-TFR-5.10-1100.
	
	\bibliographystyle{unsrtnat}

\end{document}